\documentclass[runningheads,a4paper]{llncs}

\usepackage{amssymb}
\usepackage{graphicx}

\usepackage{amsmath}
\usepackage{stfloats}
\usepackage{amsfonts}

\usepackage[misc]{ifsym}

\usepackage{url}
\urldef{\mailsa}\path|{alfred.hofmann, ursula.barth, ingrid.haas, frank.holzwarth,|
\urldef{\mailsb}\path|anna.kramer, leonie.kunz, christine.reiss, nicole.sator,|
\urldef{\mailsc}\path|erika.siebert-cole, peter.strasser, lncs}@springer.com|
\newcommand{\keywords}[1]{\par\addvspace\baselineskip
\noindent\keywordname\enspace\ignorespaces#1}

\begin{document}

\mainmatter  

\title{Watermarking Quantum Neural Networks Based on Sample Grouped and Paired Training}

\titlerunning{Watermarking QNNs Based on Sample Grouped and Paired Training}

%
%
\author{Limengnan Zhou$^1$ and Hanzhou Wu$^{2,3,*}$}
\authorrunning{Limengnan Zhou and Hanzhou Wu}
\institute{
$^1$School of Electronic and Information Engineering, University of Electronic Science and Technology of China, Zhongshan Institute, Zhongshan 528400, China\\
$^2$School of Big Data and Computer Science, Guizhou Normal University, Guiyang 550025, China\\
$^3$School of Communication and Information Engineering, Shanghai University, Shanghai 200444, China\\
$^*$\email{h.wu.phd@ieee.org}}

%
%

\maketitle

\begin{abstract}
Quantum neural networks (QNNs) leverage quantum computing to create powerful and efficient artificial intelligence models capable of solving complex problems significantly faster than traditional computers. With the fast development of quantum hardware technology, such as superconducting qubits, trapped ions, and integrated photonics, quantum computers may become reality, accelerating the applications of QNNs. However, preparing quantum circuits and optimizing parameters for QNNs require quantum hardware support, expertise, and high-quality data. How to protect intellectual property (IP) of QNNs becomes an urgent problem to be solved in the era of quantum computing. We make the first attempt towards IP protection of QNNs by watermarking. To this purpose, we collect classical clean samples and trigger ones, each of which is generated by adding a perturbation to a clean sample, associated with a label different from the ground-truth one. The host QNN, consisting of quantum encoding, quantum state transformation, and quantum measurement, is then trained from scratch with the clean samples and trigger ones, resulting in a watermarked QNN model. During training, we introduce sample grouped and paired training to ensure that the performance on the downstream task can be maintained while achieving good performance for watermark extraction. When disputes arise, by collecting a mini-set of trigger samples, the hidden watermark can be extracted by analyzing the prediction results of the target model corresponding to the trigger samples, without accessing the internal details of the target QNN model, thereby verifying the ownership of the model. Experiments have verified the superiority and applicability of this work.
\end{abstract}

\section{Introduction}
Quantum computing (QC) represents a profound paradigm shift from classical computation by leveraging quantum-mechanical phenomena such as superposition, entanglement, and interference. Recent years have witnessed significant progress in quantum hardware, with superconducting qubits (e.g., IBM's 1,121-qubit Condor processor\footnote{\scriptsize\url{https://en.wikipedia.org/wiki/IBM_Condor}}), trapped ions (e.g., IonQ's 35-qubit systems\footnote{\scriptsize\url{https://ionq.com/blog/how-we-achieved-our-2024-performance-target-of-aq-35}}), and photonic platforms (e.g., Xanadu's Borealis \cite{Madsen2022Nature}) achieving milestones in scalability and error mitigation. Despite these advances, challenges such as decoherence, high error rates, and the demand for fault-tolerant quantum error correction persist, limiting near-term applications. Concurrently, quantum machine learning and quantum-enhanced artificial intelligence (AI) algorithms have emerged as promising interdisciplinary fields. Hybrid quantum-classical models, e.g., variational quantum eigensolvers, and quantum neural networks (QNNs) demonstrate potential for exponential speedup in optimization and big data analysis. However, the current noisy intermediate-scale quantum (NISQ) era restricts practical deployments, necessitating further research into error-resilient algorithms and hardware-software co-design. The synergy between QC and AI continues to evolve, driven by theoretical breakthroughs and experimental validations.

Despite the aforementioned challenges in quantum computing, rapid advancements in quantum hardware and error mitigation techniques are steadily bridging the gap towards practical quantum computers. Once these technical hurdles are addressed, the realization of large-scale, fault-tolerant QC appears increasingly plausible. This progress underscores the critical importance of quantum AI (QAI) represented by QNNs that are quantum-enhanced algorithms providing exponential speedups for optimization, learning, and simulations, being capable of solving complex problems. Therefore, it can be foreseen that QNN models will become popular and quite important in the era of QC.

However, similar to neural network (NN) models built upon classical circuits, creating ansatz architectures and determining near-optimal parameters for QNN models require quantum hardware support, expertise, and high-quality data. As a result, the question of how to protect the intellectual property (IP) of QNNs becomes an urgent problem to be solved in the process of QAI industrialization. Fortunately, digital watermarking \cite{Wu2021TCSVT}, as a kind of information hiding technology, enables us to insert a watermark into a neural model covertly by adjusting the network without reducing utilization, which is typically referred to as \emph{model watermarking}. When disputes arise, the ownership of the target model can be reliably identified by extracting the watermark from the model, providing a very promising solution for IP protection of NN models. Therefore, it is straightforward to take into account model watermarking for protecting QNN models.

Many model watermarking methods have been reported in past years. These methods can be divided into three categories that include white-box watermarking, black-box watermarking, and box-free watermarking. White-box watermarking requires the full or partial access to the internal details of the target model during watermark extraction. It is typically realized by modulating network parameters \cite{Uchida2017ICMR}, \cite{Wang2020EI} or structures \cite{Zhao2021WIFS}. Black-box watermarking, on the other hand, does not require the access to the internals of the target model during extraction, but needs to interact with the target model. It is typically realized by injecting the watermark into a neural model through training the model with clean samples and trigger ones. These trigger samples are generated by two ways, i.e., using those samples not related to the downstream task \cite{Adi2018USENIX}, and using specifically crafted samples related to the downstream task \cite{Zhao2021ISDFS}, \cite{Zhao2024Graph}, \cite{Wang2022Symmetry}, \cite{Liu2024TDSC}. The watermark is extracted by analyzing the consistency between the assigned labels and the prediction results when inputting a set of trigger samples, for classification models. Box-free watermarking is a special case of black-box watermarking \cite{Wu2021TCSVT}, \cite{Zhang2024IH}, \cite{Zhang2022IFTP}. It extracts the watermark from the output of the target model without interacting with the model directly, which facilitates copyright protection and source tracking of AI-generated content (AIGC). This paper investigates black-box watermarking since black-box is more common than white-box.

Back to QC, a quantum is the smallest discrete unit of a physical property in quantum mechanics, which can be used for various computing tasks. The qubit is an elementary concept in quantum computing, which is the quantum version of a classical bit, capable of being in a superposition of `0' and `1' simultaneously, expressed as $\left | \psi  \right \rangle = \alpha \left | 0  \right \rangle + \beta \left | 1  \right \rangle$, where $\alpha^2 + \beta^2 = 1$. When measuring a qubit in the computational basis, it will collapse either to the state $\left | 0  \right \rangle$  or $\left | 1  \right \rangle$ with the probability $\alpha^2$ and $\beta^2$. Quantum gates manipulate qubits-quantum systems. A quantum gate is a fundamental building block of quantum circuits, analogous to classical logic gates in conventional computing. Quantum gates perform reversible operations, transforming qubit states through unitary transformations, enabling phenomena like superposition, entanglement, and interface.

A QNN contains three modules, i.e., \emph{quantum encoding}, \emph{quantum state transformation}, and \emph{quantum measurement}. Quantum encoding, e.g., maps classical input data into quantum states by amplitude encoding or angle encoding. Quantum state transformation, as the ``quantum hidden layer'' composed of trainable quantum gates, allows non-linear transformations and representations of the input quantum states. For quantum measurement, it extracts classical outputs by measuring qubits. During model training, a classical optimizer such as stochastic gradient descent (SGD) or Adam can be appropriately adjusted to optimize the tunable parameters within the quantum gates.

The proposed method marks a QNN by training the QNN from scratch with the to-be-quantum-encoded classical clean samples and trigger ones. The trigger sample is constructed by adding a perturbation (trigger) to a clean sample associated with a label different from the ground-truth. To reduce the computational complexity and maintain the performance of the QNN on its original task, we experimentally use a classical NN to reduce the dimension of the input sample and propose a sample grouped and paired strategy for training the QNN. The NN can be considered as an internal module of quantum encoding, thereby requiring the NN to be trained together with other modules. When disputes arise, by collecting a mini-set of trigger samples, the watermark can be extracted by analyzing the prediction results of the target QNN model corresponding to these trigger samples, without accessing the internal details of the target QNN model, thereby verifying the ownership of the model. Experiments show that the performance of the watermarked QNN model on the downstream task can be maintained while achieving satisfactory performance on verification.

The rest structure is organized as follows. We introduce the proposed method in Section 2, followed by experimental results and analysis in Section 3. Finally, we conclude this paper and provide discussion in Section 4.

\begin{figure*}[!t]
\centering
\includegraphics[width=\linewidth]{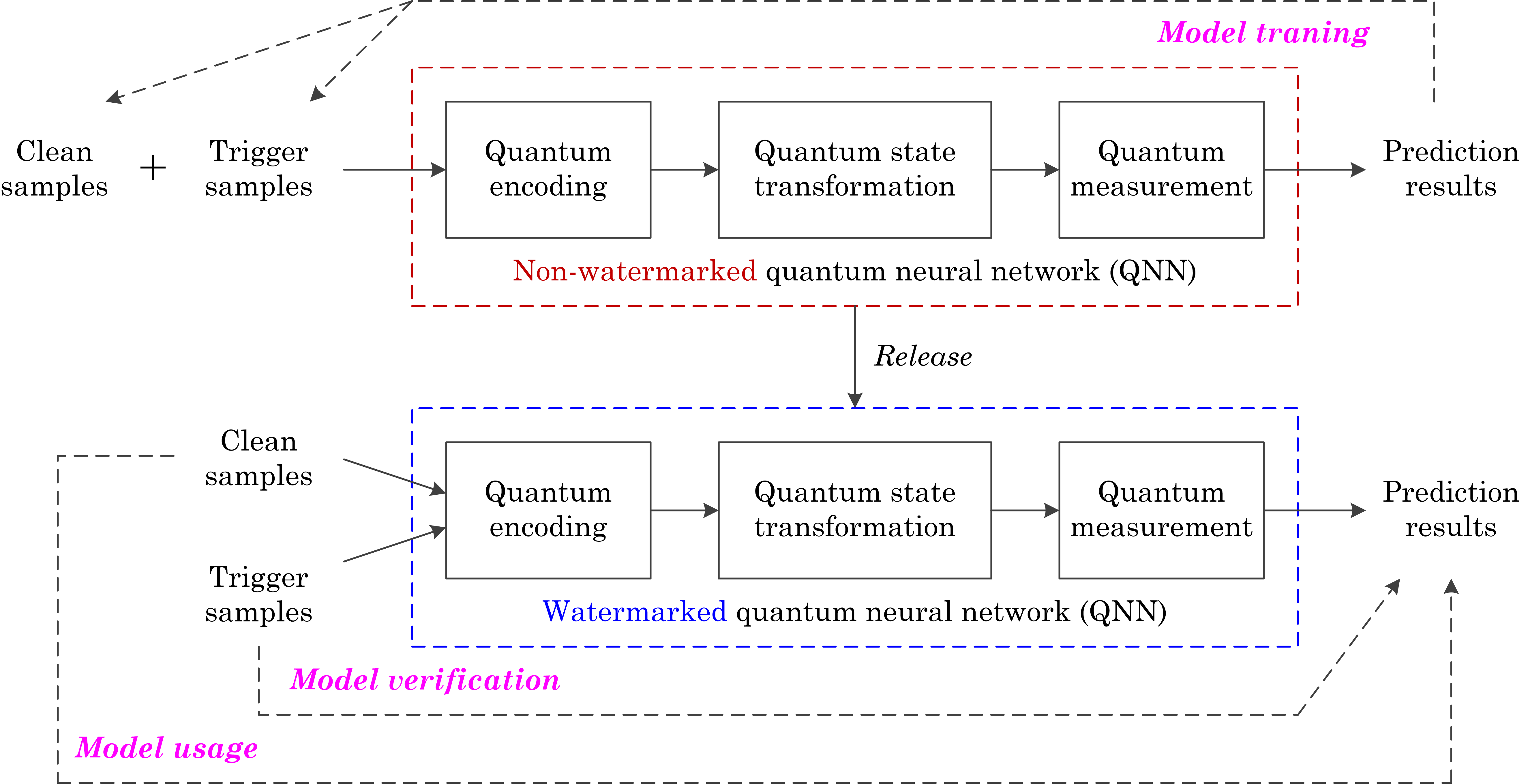}
\caption{General framework for the proposed method.}
\label{F1}
\end{figure*}

\section{Proposed Method}
Figure \ref{F1} presents the general framework including three stages, i.e., model training, model usage, and model verification. The purpose of model training is to produce the watermarked model by training the model from scratch using clean samples and trigger samples. After model training, the watermarked model is put into use, which corresponds to model usage. By analyzing the degree of consistency between the labels and the prediction results in correspondence to a set of trigger samples, the ownership of the target QNN model can be verified, which corresponds to model verification. In the following, we provide more details.

\subsection{Model Training}
Without the loss of generalization, we limit the original task of the benign QNN $\mathcal{Q}$ to be marked as image classification. This indicates that the non-marked QNN model can be generated by training with a clean image dataset $\mathcal{D} = \{(\textbf{x}_i, y_i)|1\leq i\leq |\mathcal{D}|\}$. Here, $\textbf{x}_i\in \mathbb{R}^{n\times m\times d}$ is a clean image sample, and $y_i \in \{0, 1, ..., c - 1\}$ represents the ground-truth label of $\textbf{x}_i$. It is noted that $\textbf{x}_i$ is a classical sample, which means that $\textbf{x}_i$ will be encoded before entering quantum hidden layer.

We mark $\mathcal{Q}$ by backdooring. To achieve this goal, a certain number of trigger samples $\mathcal{T} = \{(\textbf{x}_i', y_i')~|~1\leq i\leq |\mathcal{T}|\}$ should be constructed in advance. There are various ways to construct the trigger samples. We do not limit the construction of trigger samples. However, in our experiments, to verify the applicability of the proposed method, we construct trigger samples by simply adding a perturbation to clean samples. More details can be found in the experimental section.

Assuming that we have already constructed $\mathcal{D}$ and $\mathcal{T}$, $\mathcal{Q}$ will be trained from scratch with $\mathcal{D}$ and $\mathcal{T}$. During model training, our purpose is to find the near-optimal parameters so that the loss function used for fitting the clean dataset and trigger dataset can be minimal, i.e., the marked model $\mathcal{Q}_\text{marked}$ is given by
\begin{equation}
 \mathcal{Q}_\text{marked} = \underset{\mathcal{Q}^*}{\text{arg~min}}~\mathbb{E}_{(\textbf{x},y)\sim \mathcal{D}}||\mathcal{Q}(\textbf{x})-y||+\mathbb{E}_{(\textbf{x}',y')\sim \mathcal{T}}||\mathcal{Q}(\textbf{x}')-y'||.
\end{equation}
It is required that $\mathcal{Q}_\text{marked}$ not only performs well on its original task, but also exhibits satisfactory performance on watermark verification. To balance the two requirements, during model training, we adjust the order of the input samples. 

Mathematically, let $\mathcal{S}_\text{DT} = \{(\textbf{x}_1, y_1), (\textbf{x}_2, y_2), ..., (\textbf{x}_{|\mathcal{S}_\text{DT}|}, y_{|\mathcal{S}_\text{DT}|)}\}$ denote the entire sample sequence to be orderly fed into the QNN during model training. For any $1\leq i\leq |\mathcal{S}_\text{DT}|$, we have $y_i \in [0, c-1]$, where $c$ is the total classes of the samples. $\mathcal{S}_\text{DT}$ can be orderly divided into \emph{disjoint} groups (subsets), i.e.,
\begin{equation}
\mathcal{S}_\text{DT} = \mathcal{S}_{\text{D}_1} \cup \mathcal{S}_{\text{T}_1} \cup \mathcal{S}_{\text{D}_2} \cup \mathcal{S}_{\text{T}_2} \cup ... \cup \mathcal{S}_{\text{D}_q} \cup \mathcal{S}_{\text{T}_q},
\end{equation}
where $|\mathcal{S}_{\text{D}_1}| = |\mathcal{S}_{\text{D}_2}| = ... = |\mathcal{S}_{\text{D}_q}| = n_\text{D}$ and $|\mathcal{S}_{\text{T}_1}| = |\mathcal{S}_{\text{T}_2}| = ... = |\mathcal{S}_{\text{T}_q}| = n_\text{T}$.

Obviously, $q(n_\text{D} + n_\text{T}) = |\mathcal{S}_\text{DT}|$. Each $\mathcal{S}_{\text{D}_i}$, $i\in [1, q]$, contains $n_\text{D} / c$ samples with label $j\in [0,c - 1]$. Therefore, $n_\text{D}~\text{mod}~c = 0$. On the other hand, $n_\text{T}~\text{mod}~2 = 0$, implying that the samples in each $\mathcal{S}_{\text{T}_i}$, $i\in [1, q]$, are paired. It means that the samples in $\mathcal{S}_{\text{T}_i}$ can be divided into a total of $n_\text{T}/2$ disjoint pairs, each of which consists of a clean sample and its trigger version. For better understanding, the samples in $\mathcal{S}_{\text{D}_i}$, e.g., can be orderly expressed as
\begin{equation}
\{(\textbf{a}_1, 0), (\textbf{a}_2, 1), ..., (\textbf{a}_c, c-1), ..., (\textbf{a}_{n_\text{D}-c+1}, 0), (\textbf{a}_{n_\text{D}-c+2}, 1), ..., (\textbf{a}_{n_\text{D}}, c-1)\},
\end{equation}
where the first term $\textbf{a}_j$ for $j$-th sample point represents the sample data, and the second term represents the corresponding label. Similarly, the samples in $\mathcal{S}_{\text{T}_i}$, e.g., can be orderly expressed as
\begin{equation}
\{(\textbf{b}_1, l_1), (\textbf{b}_1', t), (\textbf{b}_2, l_2), (\textbf{b}_2', t), ..., (\textbf{b}_{n_\text{T}/2}, l_{n_\text{T}/2}), (\textbf{b}_{n_\text{T}/2}', t)\},
\end{equation}
where $\textbf{b}'$ is the triggered one of the clean sample $\textbf{b}$, and $t\in [0, c-1]$ is the new label assigned to all trigger samples, satisfying $l_j \neq t$ for any $j\in [1, n_\text{T}/2]$. It is possible that one may arrange the sequence by other ways. However, our experiments show that the above order already ensures satisfactory balance between watermarking and task functionality.

\subsection{Model Usage and Verification}
For model usage, it is required that the generalization of $\mathcal{Q}$ after watermarking should be well maintained, i.e., for any to-be-tested clean sample $\textbf{x}$,
\begin{equation}
\text{Pr}\{\mathcal{Q}(\textbf{x}) \neq \mathcal{Q}_\text{marked}(\textbf{x})\} < \epsilon_0 \in [0, 1],
\end{equation}
where $\epsilon_0$ is a small threshold approaching zero. Meanwhile, it is expected that $\mathcal{Q}$ and $\mathcal{Q}_\text{marked}$ exhibit significantly different prediction performance on the trigger set, i.e., for any to-be-tested trigger sample $\textbf{x}'$,
\begin{equation}
\text{Pr}\{\mathcal{Q}(\textbf{x}') = \mathcal{Q}_\text{marked}(\textbf{x}')\} < \epsilon_1 \in [0, 1],
\end{equation}
where $\epsilon_1$ is a small threshold approaching zero. From a statistical detection point of view, in order to identify the ownership of the target QNN model, we aim to maximize the difference of prediction accuracy between $\mathcal{Q}$ and $\mathcal{Q}_\text{marked}$ over the to-be-tested trigger set $\mathcal{R} = \{(\textbf{x}_i'', y_i'')~|~1\leq i\leq |\mathcal{R}|\}$, i.e.,
\begin{equation}
\left|\frac{1}{|\mathcal{R}|}\sum_{i=1}^{|\mathcal{R}|}\delta(\mathcal{Q}_\text{marked}(\textbf{x}_i''),y_i'') - \frac{1}{|\mathcal{R}|}\sum_{i=1}^{|\mathcal{R}|}\delta(\mathcal{Q}(\textbf{x}_i''),y_i'')\right| > \Delta \in [0, 1],
\end{equation}
where $\delta(x,y) = 1$ if $x = y$ otherwise $\delta(x,y) = 0$, and $\Delta$ is a \emph{large} threshold.

\begin{figure*}[!t]
\centering
\includegraphics[width=\linewidth]{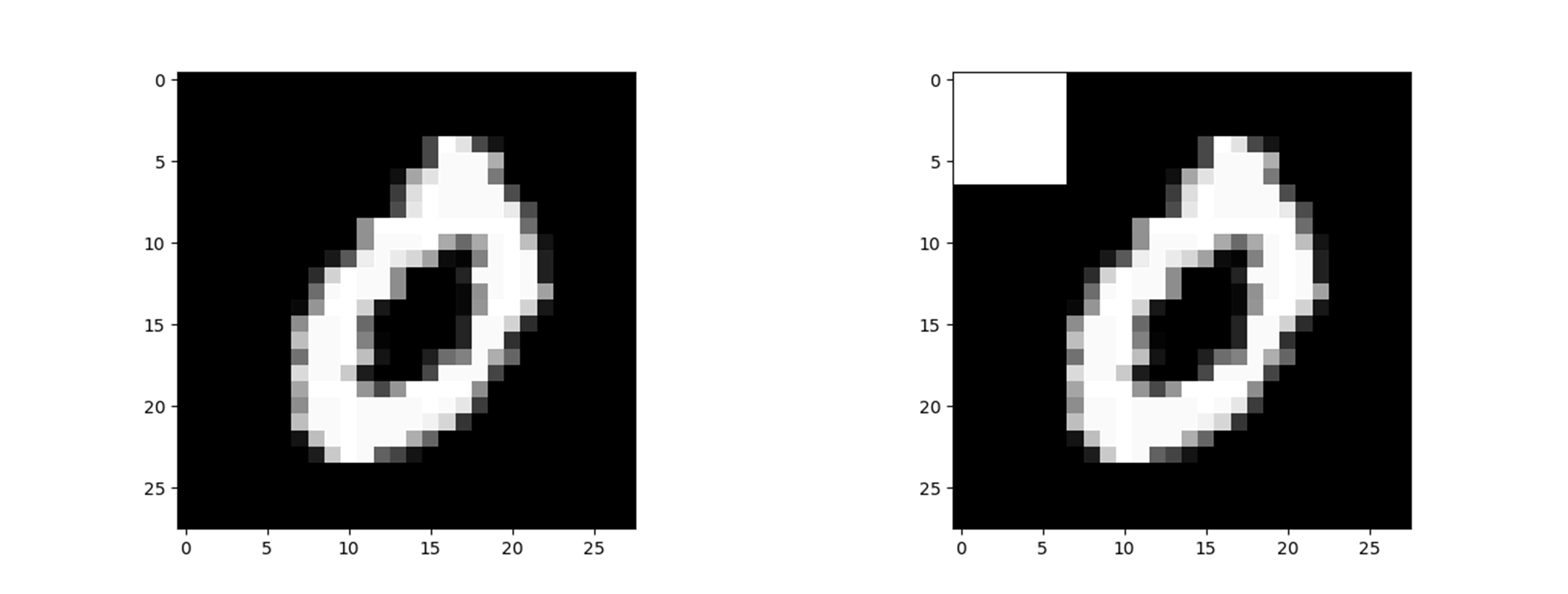}
\caption{An example of the trigger sample: clean sample (left) and trigger sample (right).}
\label{F2}
\end{figure*}

\section{Experimental Results and Analysis}
\subsection{Setup}
We conduct experiments on MNIST, a widely used benchmark in deep learning. MNIST consists of $70,000$ grayscale images of handwritten digits $(0, 1, 2, ..., 9)$, each sized $28\times 28~(=784)$ pixels. Due to the constraints in computing resources and quantum computing performance, we only select a certain number of samples for simulation. Moreover, in experiments, we only consider binary classification since QNN cannot perform very well on ordinary classification between ten digits, e.g., we have tested the performance on classifying digits `0' and `1'. We argue that reducing the number of classes to be predicted does not affect the feasibility of the proposed method since the prediction performance relies heavily on model architecture, which were not deeply studied in this paper. Our main purpose is to introduce a general framework for QNN watermarking.

Given a clean sample, its trigger version is generated by adding a visual white block to the clean sample. Figure \ref{F2} provides an example of the trigger sample, in which the size of the white block is $7\times 7$ and the value of each pixel within the white block is set to $255$. Although the design of this trigger is simple, it is enough to demonstrate the effectiveness of the proposed method. In fact, the proposed method is not limited to this trigger design, indicating that alternative approaches for trigger sample generation can also be devised.

As mentioned above, we investigate binary classification in this paper. Along this line, for each experiment, we use $100\times 2$ samples for model training. The size of the test set used for model usage is set to $100\times 2$, and that for model verification is set to $100\times 1$. We did not use the validation set since the size of the training set is not large-scale. Taking classification between digits `0' and `1' for explanation, the training set for training a clean model consists of $100$ samples for each category. To train a marked model, we use $200$ samples as well, among which $10$ samples are trigger samples and the rest are clean ones. For model usage, the test set consists of $100$ clean samples for each category. For model verification, the test set contains $100$ trigger samples. The trigger is added to the clean sample with label `0' so that the label of trigger sample is `1'.

In addition, the batch size is set to 4. The number of epochs is set to 4. We use binary cross entropy for loss determination, and Adam for loss optimization. The learning rate is set to 0.01. We use a signal GPU for acceleration.

\subsection{Model Architecture}
There are no special requirements for the architectural design of QNN. In this paper, we use hybrid quantum-classical model. A hybrid quantum-classical neural network typically consists of a classical neural network coupled with a parameterized quantum circuit in a co-processing framework. The classical network handles feature extraction, dimensionality reduction, or pre/post-processing, while the quantum component performs specialized computations such as quantum feature embedding, kernel estimation, or variational optimization. 

Along this line, we design the QNN as follows. We first apply a fully-connected neural network (FNN) to reduce the $784$-d input image into a $16$-d real-valued vector, which corresponds to quantum encoding in Figure \ref{F1}. Then, the $16$-d real-valued vector is fed into the parameterized quantum circuit to learn the quantum feature embedding, which corresponds to quantum state transformation. Finally, the quantum feature embedding is processed by another classical network (which accepts a value as input and produces a value as output) to refine predictions. This process corresponds to quantum measurement. Although the architecture designed in this paper is simple and not a purely quantum network architecture, it demonstrates the effectiveness of the proposed method. We use \emph{Qiskit}\footnote{\url{https://www.ibm.com/quantum/qiskit}}, which is an open-source, Python-based, high-performance software stack for quantum computing, originally developed by IBM, for simulation. 

\begin{figure*}[!t]
\centering
\includegraphics[width=\linewidth]{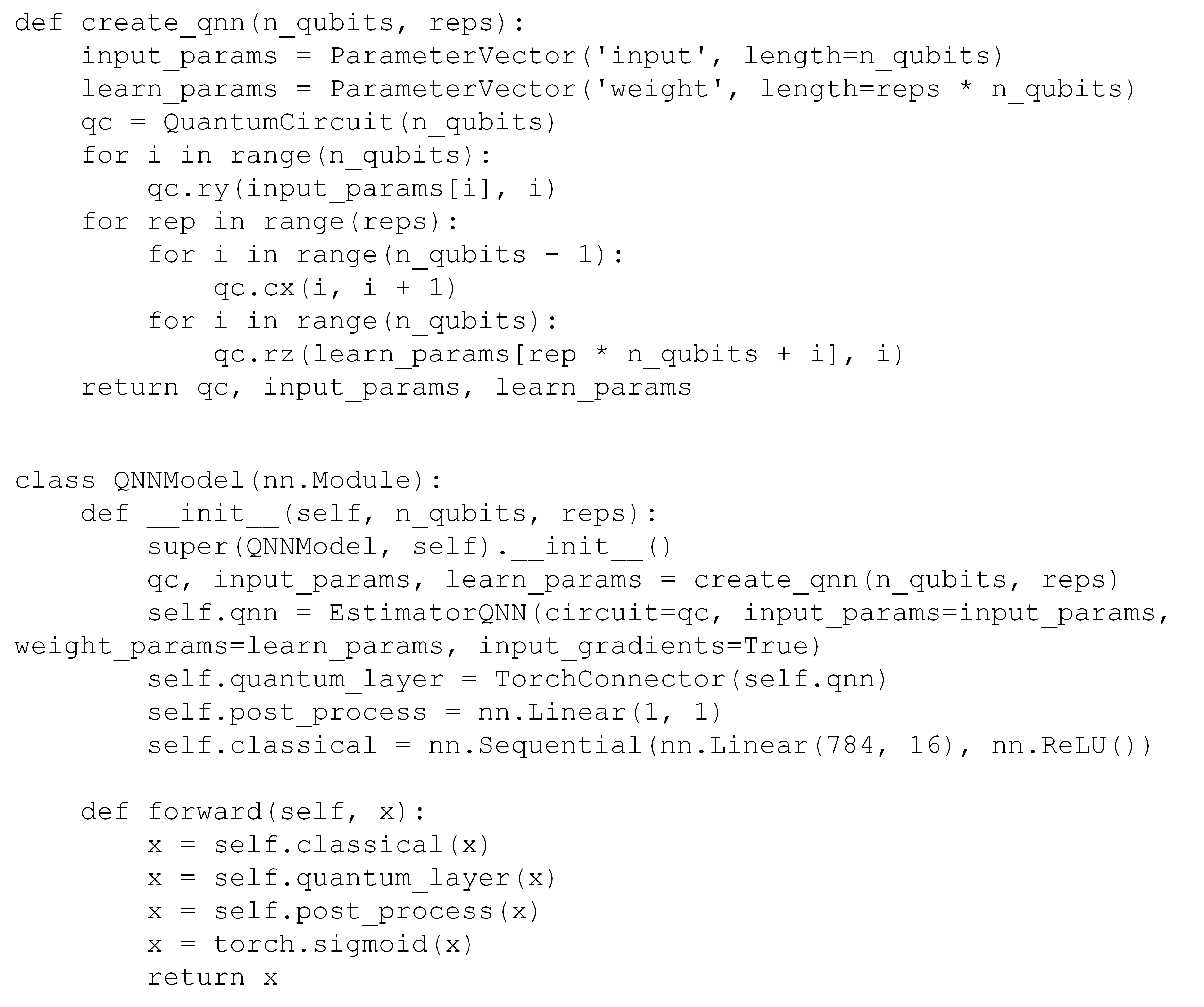}
\caption{Source code for the model architecture (with Python, PyTorch, and Qiskit).}
\label{F3}
\end{figure*}

Figure \ref{F3} shows the Python code for building the QNN model. As shown in Figure \ref{F3}, the total number of trainable parameters for quantum state transformation is $\texttt{reps}\ast \texttt{n\_qubits}$. In our experiments, we set $\texttt{reps} = 2$ and $\texttt{n\_qubits} = 16$. Instead of applying amplitude encoding or angle encoding to the output of the classical neural network, we here directly fed the classical output into the hidden quantum layer since the classical network and the hidden quantum layer will be trained together so that the classical output could adapt to the hidden input. 

It can also be seen that quantum state transformation uses CX (Controlled-X) Gate, RY (Rotation-Y) Gate and RZ (Rotation-Z) Gate. The CX gate is a two-qubit entangling gate that flips the target qubit (applies an X gate) if and only if the control qubit is in state $\left | 1  \right \rangle$. It is fundamental for creating quantum correlations (entanglement) and forms the basis of many quantum algorithms. The RY gate performs a single-qubit rotation around the Y-axis of the Bloch sphere by angle $\theta$. It is widely used for preparing superpositions and variational quantum circuits. The RZ gate rotates a qubit around the Z-axis by $\theta$, introducing a phase difference between the $\left | 0  \right \rangle$ and $\left | 1  \right \rangle$ states without changing their probabilities, which is critical for quantum phase manipulation.

\begin{table*}[!t]
\caption{Performance evaluation of model watermarking on binary classification.\label{tab:table1}}
\centering
\begin{tabular}{c|c||c||c|c}
\hline\hline
Category 1 & Category 2 & \#\{trigger set\} & Acc. (clean set)  & Acc. (trigger set) \\ \hline
0 & 1 & 0 & 86\% & 23\% \\ \hline
0 & 1 & 10 & 83.5\% & 97\% \\ \hline\hline
2 & 3 & 0 & 81.5\% & 37\% \\ \hline
2 & 3 & 10 & 79\% & 98\% \\ \hline\hline
4 & 5 & 0 & 85\% & 37\% \\ \hline
4 & 5 & 10 & 83.5\% & 82\% \\ \hline\hline
6 & 7 & 0 & 77\% & 32\% \\ \hline
6 & 7 & 10 & 75\% & 79\% \\ \hline\hline
8 & 9 & 0 & 93\% & 40\% \\ \hline
8 & 9 & 10 & 90.5\% & 80\% \\ \hline\hline
\end{tabular}
\end{table*}

\begin{table*}[!t]
\caption{Performance evaluation after increasing the size of trigger set.\label{tab:table2}}
\centering
\begin{tabular}{c|c||c||c|c}
\hline\hline
Category 1 & Category 2 & \#\{trigger set\} & Acc. (clean set)  & Acc. (trigger set) \\ \hline
4 & 5 & 10 & 83.5\% & 82\% \\ \hline
4 & 5 & 20 & 80.5\% & 100\% \\ \hline\hline
6 & 7 & 10 & 75\% & 79\% \\ \hline
6 & 7 & 20 & 72.5\% & 100\% \\ \hline\hline
8 & 9 & 10 & 90.5\% & 80\% \\ \hline
8 & 9 & 20 & 89\% & 100\% \\ \hline\hline
\end{tabular}
\end{table*}

\subsection{Results}
We are ready to report our experimental results. Table \ref{tab:table1} shows the performance on model watermarking, in which different digit pairs are evaluated. When the size of the trigger set is zero, it means that we are concerned about the performance of the non-watermarked model on the original classification task. We set the size of the trigger set to $10$ by default to mark the QNN. It is noted that, for those watermarked models, they are trained with the proposed strategy, while the non-watermarked ones are trained with random sample permutation. In addition, we generate the trigger samples by adding the trigger pattern to clean samples belonging to Category 1. We set $n_\text{T} = 2$ and $|\mathcal{S}_\text{DT}| = 200$ by default.

As shown in Table \ref{tab:table1}, different parameter settings generally result in different performance. Overall, given the baseline, after watermarking, the performance on the original task tends to decline, which is reasonable since more tasks require more representation abilities of the model, thereby reducing the generalization ability on a specific task. However, the performance degradation on the original task can be kept within a relatively low level. Moreover, the difference of the classification accuracy on the trigger set between the non-watermarked model and the watermarked model is significant for all five cases, indicating that the proposed method enables us to reliably embed and extract the watermark. This shows that the proposed method is effective for IP protection of QNN models.

It can be observed from Table \ref{tab:table1} that some digit pairs give relatively smaller classification accuracy values on the trigger set, which should be affected by the size of the trigger set. To confirm this, we increase the size of the trigger set. As shown in Table \ref{tab:table2}, the trigger samples can be perfectly classified, at the expense of slight performance decline on the original task. Therefore, in applications, it is suggested to properly adjust the ratio between the size of the clean samples and the size of the trigger samples so that a good performance trade-off between the original task and the watermarking task can be achieved.

We have also evaluated the performance using the proposed training strategy and random ordering strategy (i.e., the training samples are randomly ordered). Simulation results indicate that the proposed method gives better watermarking performance and has better stability than random ordering in terms of watermark extraction, i.e., random orders give significant different performance, e.g., some random orders cause watermark extraction to fail completely in our experiments, while this can be largely avoided with the proposed method.

\section{Conclusion and Discussion}
In the era of rapid advancements in QC and AI, protecting intellectual property and ensuring model integrity have become paramount. Watermarking quantum neural network models is a critical step to safeguard innovation, prevent unauthorized use, and maintain accountability in AI deployments. In this paper, we present an effective watermarking framework for QNN models. The main idea is to use trigger samples to train the clean model so that the resulting watermarked model enables us to verify its ownership by analyzing the prediction results given the input samples. In order to balance the performance of the model on the original task and the watermarking task, we introduce sample grouped and paired training strategy for updating the trainable quantum parameters. Experimental results verify its feasibility and effectiveness.

The rapid progress in QC suggests that practical, large-scale quantum machines will likely become a reality in the foreseeable future. As these systems approach computational supremacy, they will inevitably power the next generation of artificial intelligence, enabling breakthroughs in optimization, cryptography, and machine learning. However, this transformative potential also introduces unprecedented security risks. Quantum-driven AI models, with their enhanced capabilities, could be exploited for malicious intents. It is now the time to develop safeguards, before quantum AI becomes mainstream, ensuring a secure and trustworthy foundation for the future of intelligent systems. We hope this preliminary exploration could motivate researchers focused on information hiding to pay more attention on IP protection of quantum AI models.

\section*{Acknowledgement}
This work was partly supported by the Basic Research Program for Natural Science of Guizhou Province under Grant Number QIANKEHEJICHU-ZD[2025]-043, and also partly supported by the Guangdong Basic and Applied Basic Research Foundation under Grant Number 2023A1515010815.

\end{document}